# R-M INTERACTIONS IN $R_2BaMO_5$ (R=Y or Gd AND M=Cu or Zn)

G.F. Goya and R. C. Mercader

*Departamento de Física, Universidad Nacional de La Plata,*
*1900 La Plata, Argentina*

L.B. Steren, R.D. Sánchez, M.T. Causa, and M. Tovar

*Instituto Balseiro and Centro Atómico Bariloche,*
*8400 Bariloche, Argentina*

## ABSTRACT

$R_2BaMO_5$ (R = Gd, Y and M = Cu, Zn) oxides have been studied by specific heat, dc magnetic susceptibility, and electron-paramagnetic-resonance (EPR). For one member of the series without magnetic moment at M, namely $Gd_2BaZnO_5$, measurements reveal long range antiferromagnetic order at $T_N(Gd_2BaZnO_5) = 2.3\pm0.1K$, much lower than the Curie-Weiss temperature of $\Theta=15.9\pm0.3K$. This indicates the existence of competing interactions that introduce a large degree of magnetic frustration in the system. For $Y_2BaCuO_5$ the Cu-Cu interactions are responsible for the broad maximum in the magnetic contribution to the specific heat centered at $18.5\pm0.01K$ that stretches beyond the instrumental limit of 25K. The strong Cu-Cu interactions also present in $Gd_2BaCuO_5$, combined with the Gd-Cu interaction, polarize the Gd sublattice giving measurable contributions to the specific heat in the same temperature range. In addition, they broaden the Gd EPR line and saturate its integrated intensity. The ordering temperature of Gd ions is raised to $T_N(Gd_2BaCuO_5)=12.0\pm0.1K$.

PACS numbers: 75.30.-m; 75.30.Et; 75.50.Ee

I. INTRODUCTION

Magnetic interactions in the $R_2BaMO_5$ family of ternary oxides (R=rare earth or Yttrium and M=3d transition metal) have been analyzed in recent papers[1-3]. Attention to these systems has been driven by the different crystalline and magnetic structures of the members that change according to the nature of the R and M elements, providing in some cases examples of low-dimensional magnetism[3-5].

Additional interest rises because structural and electron difference density measurements[6] have shown that the Cu at Cu(2) sites of the high-$T_c$ superconductor $YBa_2Cu_3O_{7-\delta}$, which mediates the superconducting mechanism, has similar site geometry and electronic configuration as Cu in the insulating "green phase" $Y_2BaCuO_5$.

Specific heat and magnetic susceptibility measurements reported for most $R_2BaCuO_5$ compounds, when R is a magnetic ion, have been interpreted in terms of two separate magnetic transitions[7,8] associated with the magnetic ordering of the R and Cu sublattices ($T_{N1}$ and $T_{N2}$ respectively). These temperatures vary with the different R ions involved, laying within the range $T_{N1} \_ 12K$ and $15K \_ T_{N2} \_ 23K$. On the other hand, recent neutron diffraction experiments suggest[9] that magnetic order appears simultaneously in the R and Cu sublattices. Below $T_{N2}$, the magnetic ordering of the Cu spins partially polarizes the R spins due to the magnetic R-Cu coupling. Further cooling causes saturation of the R moments at a lower temperature $T_{N1}$. Although the microscopic interaction between the two magnetic subsystems has not been fully analyzed, it has been suggested[2] that the presence of Cu at M sites is necessary to induce the magnetic ordering at the R site.

In the presence of two interacting magnetic subsystems, namely the R and M ions, it seems necessary to investigate first the R-R and M-M interactions in order to clarify the nature of the coupling of the R and M subsystems. To clarify this question, we have performed magnetic susceptibility, specific heat and EPR measurements on the same set of samples of the isomorphous compounds: $Gd_2BaZnO_5$, with magnetic Gd ions at R sites and non-magnetic Zn at M sites, Y2BaCuO5, with non-magnetic Y and magnetic Cu, and Gd2BaCuO5 with both magnetic species present simultaneously.

| Compound | $T_N$ (K) | $\mu_{eff}$ ($\mu_B$) | $\Theta$ (K) | $\Delta H_{PP}$ (mT) | $g_{eff}$ [‡] |
|---|---|---|---|---|---|
| $Y_2BaCuO_5$ | † | 1.87(1) | -43(4) | 245(9) | 2.126(3) |
| $Gd_2BaZnO_5$ | 2.3(1) | 7.99(1) | -15.9(3) | 2600(100) | 1.99(2) |
| $Gd_2BaCuO_5$ | 12.0(1) | 16.48 | 23.2(3) | 1090(30) | 2.017(3) |

**Table I.** *Néel temperature $T_N$, magnetic moment $\mu_{eff}$, Weiss temperature $\Theta$, EPR peak-to-peak linewidth $\Delta H_{PP}$ and gyromagnetic factor $g_{eff}$ values for the powder samples. †: No ordering temperature is observed from the susceptibility data. ‡ : This value correspond to policrystalline samples and is therefore an average over the components of the g-tensor.*

II. EXPERIMENTAL

Samples of $R_2BaMO_5$, for R=Y, Gd and M=Zn, Cu, were prepared by standard solid state reaction of $R_2O_3$ (99.99%), $BaCO_3$ (99.99%), and MO(99.999%) powders, mixed in stoichiometric amounts. The homogenized mixture was first fired in air at 500 °C, then reground and reheated three times at 900 °C for 24 hours. Powder X-ray diffraction characterization was performed using Cu-Kα radiation. Data were refined by Rietveld profile analysis[10] showing in all cases a single phase in the Pnma group, according to previous structural data on these systems[11,12]. DC magnetic susceptibility was measured with a SQUID magnetometer in the temperature range 1.7 - 300K. In all cases, susceptibilities were corrected for core diamagnetism from the ions. EPR data were obtained with a Bruker ESP300 spectrometer for X-band (ν ≈ 9.4 GHz) between 2.5 and 300K, and at room temperature for Q-band (ν ≈ 35 GHz). The powder EPR spectra were processed by comparison with simulated spectra. We have included the necessary corrections[13] to superposition of resonant lines at positive and negative fields when the peak-to-peak linewidth, $\Delta H_{PP}$, was comparable to the resonance field $H_R$. Specific heat measurements were obtained by a pulsed method in a semiadiabatic calorimeter up to 25K. In order to subtract the non-magnetic contributions to $C_P$ in $Y_2BaCuO_5$, we performed specific heat measurements on the isomorphic compound $Y_2BaZnO_5$, including corrections for the small difference in molar mass.

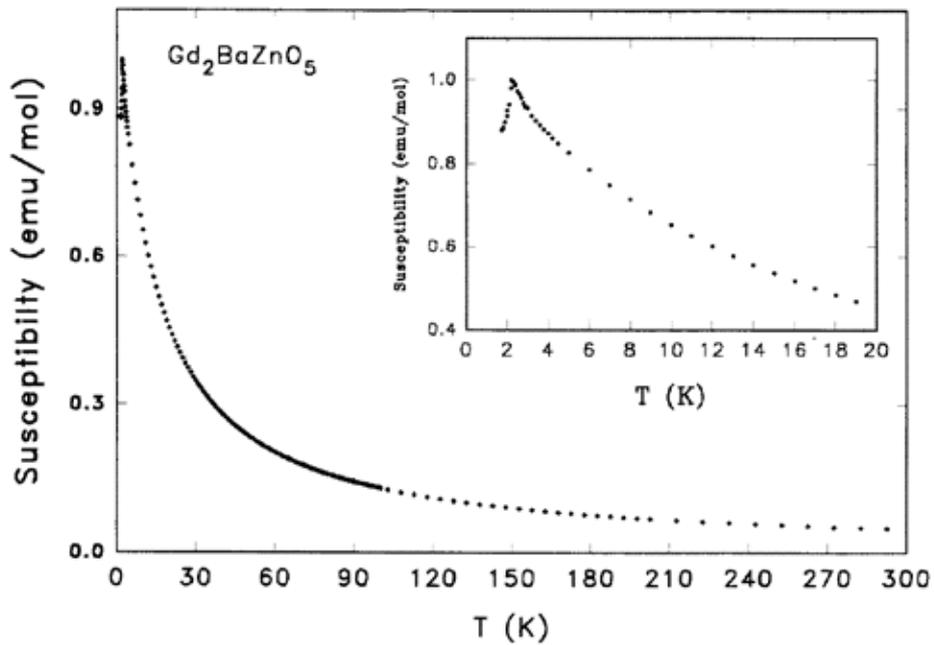

*Figure 1.* Magnetic susceptibility for Gd2BaZnO5 versus *T* , measured with *H* = 50 mT. The inset shows the peak at *TN* = 2:3(1) K.

## III. RESULTS AND DISCUSSION

The magnetic susceptibility of $Gd_2BaZnO_5$ was found to be independent of the applied field H. Figure 1 shows the data measured between 1.8K and 300K, with H=50mT. The high temperature data (100K-300K) were fitted to a Curie-Weiss law, $\chi(T)=C/(T-\Theta)$, with C = 8.00±0.03emu-K/mol and $\Theta$=-15.9±0.3K (see Table I). This value of C yields a calculated effective moment, $\mu_{eff}$=7.99±0.01$\mu_B$, close to the expected value for free $Gd^{+3}$ ions (S=$7/2$; $g_{eff}$=1.991). Since the effect of the crystalline field on the ground S-state is usually very small, it gives no measurable contributions to $\chi(T)$ in the temperature range of our experiment. The negative value of $\Theta$ indicates antiferromagnetic (AF) interactions between Gd ions. A peak at $T_N(Gd_2BaZnO_5)$ = 2.3±0.1K is clearly noticed, indicating long-range ordering of the Gd ions.

Susceptibility measurements on $Y_2BaCuO_5$ show a paramagnetic regime at high temperatures (100K - 300K), following a Curie-Weiss law with $\mu_{eff}$= 1.87±0.01$\mu_B$ and $\Theta$=-43±4K (Table I), in agreement with previous results[14,15]. Below 55K, dc susceptibility deviates from paramagnetic behaviour and exhibits a broad maximum at $T(\chi_{max}) \approx$ 32K, well above the ordering temperature determined from neutron diffraction experiments[16], $T_N(Y_2BaCuO_5)$ = 16.5K. Our data display non-linearity of the M vs. H curves in the region T_15K, in agreement with this ordering temperature.

DC susceptibility data of $Gd_2BaCuO_5$ display a peak at $T_N(Gd_2BaCuO_5)$ = 12.0±0.1K, and follows a Curie-Weiss law[8,17] at high temperatures (100K-300K). The fitted parameters are shown in Table I. The value of the Curie constant corresponds with the sum of contributions from Gd and Cu ions. The $\chi^{-1}(T)$ vs. T curves show deviations from the Curie-Weiss law for T_35K. This kind of deviation is not observed in $Gd_2BaZnO_5$ down to

T≈3K.

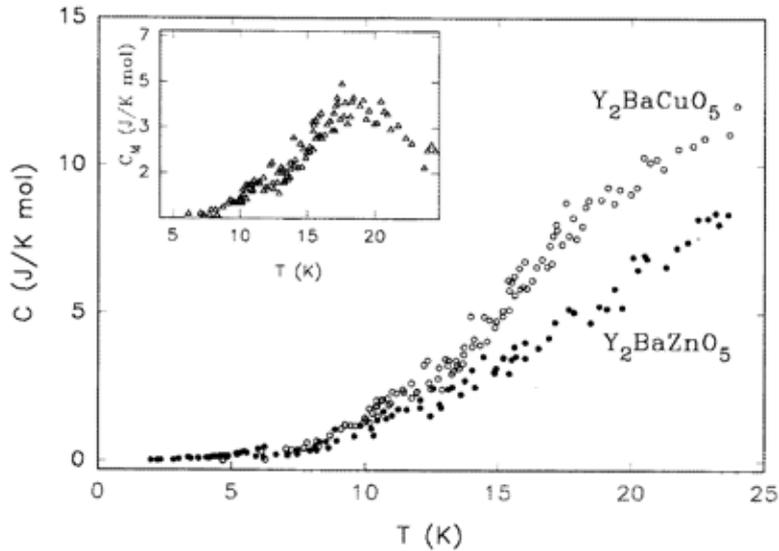

*Figure 2.* *Specific heat of magnetic $Y_2BaCuO_5$ and non-magnetic $Y_2BaZnO_5$. The inset displays the Cu magnetic contribution, as obtained from the $C_p(Y_2BaCuO_5)$ - $C_p(Y_2BaZnO_5)$ difference.*

The specific heat data for $Y_2BaCuO_5$ and non-magnetic $Y_2BaZnO_5$ are shown in Fig. 2. The magnetic contribution estimated from the difference of the two curves presents a broad maximum around $T_{max} \approx 18.5K$. This maximum corresponds to the change of slope and the rounded anomaly reported in Refs.7 and 18, respectively. Non-negligible contributions are observed well above $T_{max}$, up to our experimental limit (T ≈ 25K). Only 35% of the magnetic entropy is removed below $T_{max}$ and 57% below 25K.

Specific heat measurements in $Gd_2BaCuO_5$, shown in Fig. 3, present a λ-type anomaly at $T_{N1}(Gd_2BaCuO_5)$ = 11.8±0.1K, in coincidence with the dc susceptibility peak. The entropy removed below $T_{N1}$ corresponds to about 69% of the value expected for S = 7/2, $\Delta S(Gd) = 2R\ln 8$. Above $T_{N1}$ the magnetic contribution is still significant. The magnetic entropy removed up to the small bump observed at $T_b = 18.0\pm0.5$K, $\Delta S = 3.6R$, is about 87% of $\Delta S(Gd)$. If the magnetic entropy associated with the Cu moments is included, the entropy removed is about 75% of the total value.

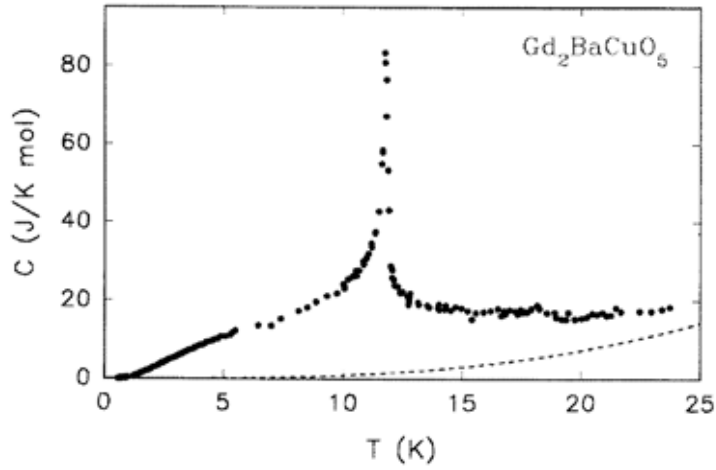

*Figure 3. Specific heat of $Gd_2BaCuO_5$: The dashed line ( – – –) represents the lattice contribution.*

Figure 4 shows the X-band EPR spectra for $Y_2BaCuO_5$, $Gd_2BaCuO_5$ and $Gd_2BaZnO_5$, measured at room temperature. For $Y_2BaCuO_5$ the Cu ions give a resonance line with a structure that reflects the g-factor anisotropy. At Q-band this structure is completely resolved in three components that correspond to the rhombic symmetry of the g-tensor[19]. The fit of the powder spectrum with computer simulated profiles gives $g_x = 2.229\pm0.003$, $g_y = 2.096\pm0.003$, and $g_z = 2.054\pm0.003$, in close agreement with previous results[19]. The intrinsic linewidth was also obtained from the fit. At room temperature we determined for both X- and Q-band $\Delta H_{pp}(300K) = 8.0\pm0.5$ mT, which is slightly larger than the values found in single crystals[20]. The observed full width, $\Delta H_{pp}(300K)=24.5\pm0.9$ mT for X-band and 97.5±0.9 mT for Q-band, is much larger than the intrinsic linewidth due to the anisotropy of the g-factor.

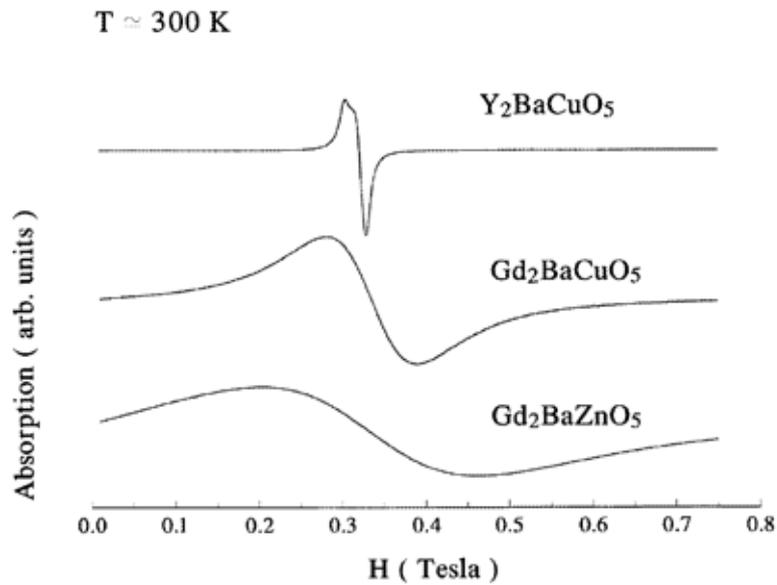

*Figure 4.* Room-temperature X-band EPR spectra of Y2BaCuO5, Gd2BaCuO5 and Gd2BaZnO5.

To examine the exchange interaction near the ordering temperature we have determined the variation of the EPR linewidth and the integrated intensity vs. temperature, shown in Fig.5. The integrated intensity reaches a maximum around 20K, above to the ordering temperature of the Cu ions[16]. Below $T_N$ the intensity of the resonance line is expected to decrease as a consequence of the development of a field dependent energy gap in frequency-field diagram. For antiferromagnetically coupled moments this gap is usually anisotropic[21] and for a powder sample antiferromagnetic resonance (AFMR) would only be observed for a fraction of the microcrystals.

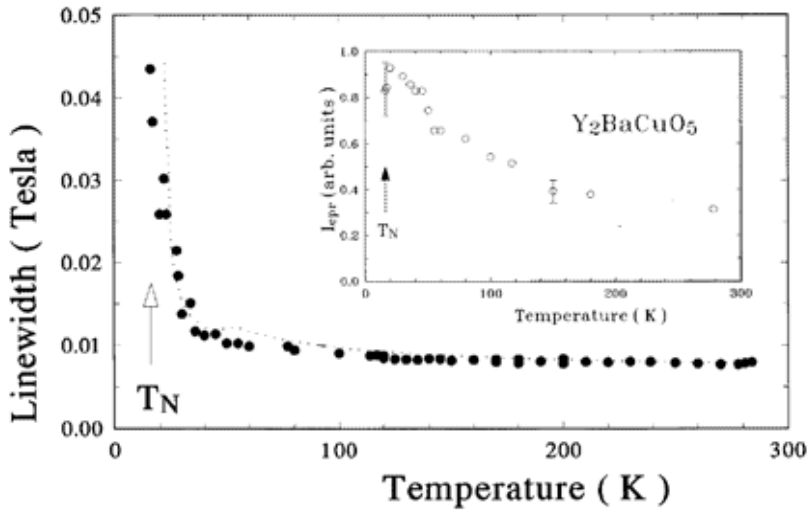

**Figure 5.** Temperature dependence of $\Delta H_{pp}$ for $Y_2BaCuO_5$: _ _ _ _ _ _, $\Delta H_{pp}(T) = \Delta H_{pp}(\infty)/T\chi(T)$. The inset shows the integrated intensity versus T.

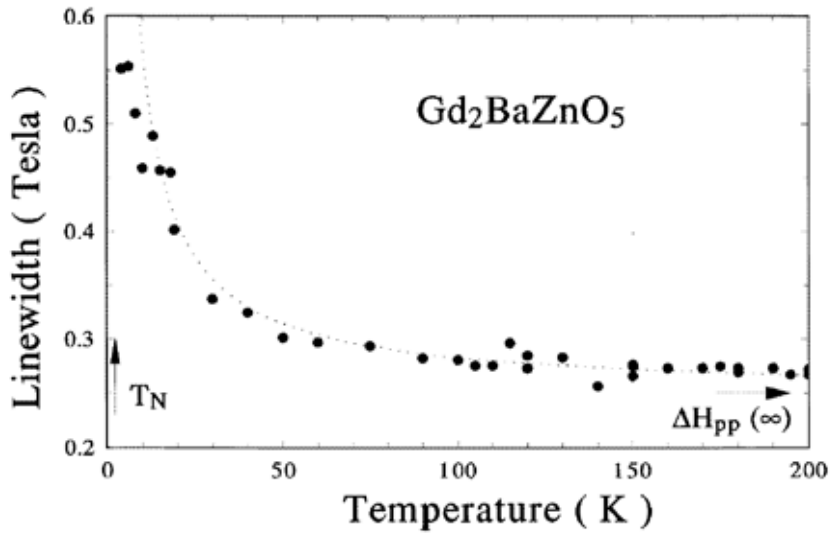

**Figure 6.** Temperature dependence of $\Delta H_{pp}$ for $Gd_2BaZnO_5$: _ _ _ _ _ _, $\Delta H_{pp}(T) = \Delta H_{pp}(\infty)/T\chi(T)$.

Thus, the maximum at 20K indicates the presence of short range AF order above $T_N$. Notice that the broad peak reported at around 54K in Ref.17 may be a consequence of estimating the intensities as the peak-to-peak amplitude multiplied by the square of the measured linewidth[22]. In this case the procedure would be incorrect, because the observed peak-to-peak linewidth is due to the g-anisotropy and numerical integration is required.

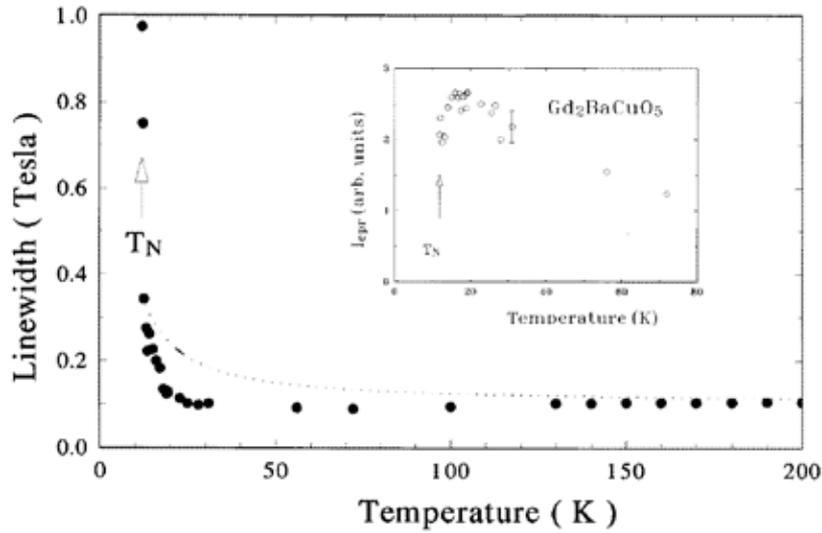

**Figure 7.** Temperature dependence of $\Delta H_{pp}$ for Gd2BaCuO5 (dashed line), $\Delta H_{pp}(T) = \Delta H_{pp}(\infty)/T\chi(T)$. The integrated intensity versus T of the EPR line is shown in the inset.

For $Gd_2BaZnO_5$, EPR measurements show a single broad line associated with the Gd ions, centered at g = 1.99(2) and with a room temperature linewidth $\Delta H_{pp}(300K) = 0.26(1)$ T. Figure 6 shows that the linewidth increases slowly with decreasing temperature, following approximately an expression $\Delta H_{pp}(T) \propto \Delta H_{pp}(\infty)/T\chi(T)$ in accordance with the behaviour found in other magnetic compounds[22,23]. This expression gives a constant linewidth for paramagnetic Curie-law systems. In the case of antiferromagnetic compounds, where the susceptibility respond to a Curie-Weiss law, the expression $\Delta H \propto (1+\Theta/T)$ predicts a monotonic broadening of the linewidth when the temperature decreases. This broadening is observed in an extended region of T ( 3K _ T _ 200K ) because the system orders at $T_N \approx \Theta/7$ due to its high degree of frustration.

In the case of $Gd_2BaCuO_5$, where two magnetic species are present, the EPR line is expected to correspond mainly to the Gd$^{+3}$ moments, due to their much larger dc susceptibility. A single line, with no structure, is observed in this case. A value of g = 2.017±0.003 was determined from Q-band data. The linewidth, $\Delta H_{pp}$(300K) = 109±3 mT for X-band and 124±3 mT for Q-band, is about three times smaller than in the case of Gd$_2$BaZnO$_5$ and reflects the exchange coupling of the Cu and Gd sublattices. The Gd$_2$BaCuO$_5$ linewidth, unlike that of Gd$_2$BaZnO$_5$, decreases slightly with decreasing T reaching a shallow minimum at $\Delta H_{pp}$(70K) ≈ 92±3 mT, as shown in Fig. 7. The fast increase in $\Delta H_{pp}$ due to each interaction is observed below ≈30K, in agreement with Ref.17. This linewidth increase produces a rapid diminution of the peak-to-peak line amplitude and the signal is not longer observed below T ≈ 12K. The integrated intensity, in turn, increases up to 20K, where it reaches its maximum value.

The crystalline structure of *Pnma*-R$_2$BaMO$_5$ consists of distorted monocapped trigonal prisms of RO$_7$, which form a three-dimensional framework. Isolated MO$_5$ pyramids are located in cavities delimited by the RO$_7$ framework[12,6]. This complex structure makes possible the existence of several exchange pathways between magnetic ions, even in Gd$_2$BaZnO$_5$ and Y$_2$BaCuO$_5$, where only one magnetic species is present. This observation may explain the high value of $\Theta/T_N$ for these compounds, indicative of a strong degree of magnetic frustration. On the other side, the presence of Cu moments, with a large Cu-Cu antiferromagnetic interaction (see Table I) reduces the $\Theta/T_N$ ratio from 7 in the case of Gd$_2$BaZnO$_5$ to 2 in Gd$_2$BaCuO$_5$.

Based on the neutron diffraction results for other $R_2BaCuO_5$ compounds ( R = Y, Dy, Ho, Er, Tm and Yb)[9,16] we can suggest the following picture for $Gd_2BaCuO_5$: at a temperature around $T_N(Y_2BaCuO_5)$ = 16.5K, the Cu moments are expected to order AF. The saturation of the EPR intensity at ≈20K, and the small bump at 18K in the specific heat of $Gd_2BaCuO_5$ might signal this temperature. The Cu ordering would in turn polarize the Gd moments, as they do in other $R_2BaCuO_5$ and $R_2BaNiO_5$ compounds [5,9]. As the temperature of the coupled system is lowered, the Gd polarization increases. The large peak of $C_P$ at $T_{N1}(Gd_2BaCuO_5)$ = 12K would indicate the temperature of maximum increasing of the Gd sublattice magnetization with T. The fact that the entropy removed between 12 and 18K, $\Delta S/R = 0.8\pm0.1$, is much larger than the value measured in $Y_2BaCuO_5$ for the ordering of Cu ions up to the same temperature, also suggests that the Gd ions start to order above $T_{N1}(Gd_2BaCuO_5)$, thus contributing to the total specific heat. It seems therefore possible that, in $Gd_2BaCuO_5$, the exchange coupling between Gd and Cu moments helps to overcome the magnetic frustration of the Gd subsystem (observed in $Gd_2BaZnO_5$) at a much higher temperature than $T_N(Gd_2BaZnO_5)$ = 2.3K. The likelihood of this description is also supported by the saturation of the Gd EPR intensity at 20 K, well above the ordering temperature.

In summary, we have observed long-range magnetic order for $Gd_2BaZnO_5$, a compound of the *Pnma*-$R_2BaMO_5$ family with no magnetic moments at the M sites. Contrary to previous indications[2], this result establishes that the existence of a magnetic M sublattice is not necessary to induce magnetic ordering of the R lattice. Nonetheless, the relevance of the Gd-Cu interactions can be noticed since they rise the ordering temperature by a factor of 6, from $T_N(Gd_2BaZnO_5)$ = 2.3K to $T_N(Gd_2BaCuO_5)$ = 12.0K, in spite of the similarity between Gd-Gd interactions suggested by the corresponding Curie-Weiss temperatures. Moreover, the EPR results indicate that magnetic order in this compound extends well above 12K. A large tail in the specific heat and a small feature at 18K suggest that the Néel temperature of the coupled Gd-Cu system may be higher than 12K, temperature where the specific heat and the magnetic susceptibility reach their maximum value. Additionally, our data show that short-range order in $Y_2BaCuO_5$ extends above the Néel temperature $T_N$=16.5K determined from neutron scattering experiments, although not as high as the 54K estimated previously[17].

**ACKNOWLEDGEMENTS**

We acknowledge partial support from Consejo de Investigaciones Científicas y Técnicas through the LANAIS and TENAES programs, and PID 92 and from the Commission of the European Communities DGXII, contract CI1*CT92-0087. G.F.G. also acknowledges support from CONICET and Comisión Nacional de Energía Atómica during his stay in Bariloche.